\documentclass[a4paper, BackFigs,11pt]{article}
\usepackage{amsfonts}
\usepackage{amsmath}
\usepackage{amssymb}
\usepackage{graphicx}%
\usepackage{color}
\usepackage{float}
\usepackage[numbers,sort&compress]{natbib}

\title{Broadband ultra-high transmission of terahertz radiation through monolayer MoS$_{2}$}

\author{\small Xue-Yong Deng$^{1,2}$, Xin-Hua Deng$^{3}$, Fu-Hai Su$^{4}$, Nian-Hua Liu$^{1,3}$, Jiang-Tao Liu$^{1,3*}$\\  \footnotesize
$^{1}$Nanoscale Science and Technology  Laboratory,  Institute for Advanced Study, \\
 \footnotesize  Nanchang University, Nanchang 330031, China \\  \footnotesize  $^{2}$School of Materials Science and Engineering,
 Nanchang University, Nanchang 330031, China\\
 \footnotesize $^{3}$
Department of Physics, Nanchang University,
Nanchang
330031, China\\
\footnotesize
 $^{4}$Key Laboratory of Materials Physics, Institute of Solid State Physics,\\  \footnotesize
Chinese Academy of Sciences, Hefei 230031, People¡¯s Republic of China\\
 \footnotesize   $^{*}$Email: jtliu@semi.ac.cn
 }

\begin{document}

\maketitle
\begin{abstract}
In this study, terahertz (THz) absorption and transmission of monolayer MoS$_{2}$ was calculated under different carrier concentrations. Results showed that the THz absorption of monolayer MoS$_{2}$ is very small even under high carrier concentrations and large incident angle. Equivalent loss of the THz absorption is the total sum of reflection and absorption that is one to three grades lower than that of graphene. The monolayer MoS$_{2}$ transmission is much larger than that of the traditional GaAs and InAs two-dimensional electron gas. The field-effect tubular structure formed by the monolayer MoS$_{2}$-insulation-layer-graphene is investigated. In this structure the THz absorption of graphene to reach saturation under low voltage. Meantime, the maximum THz absorption of monolayer MoS$_{2}$ was limited to approximately 5\%. Thus, monolayer MoS$_{2}$ is a kind of ideal  THz Transparent Electrodes.
\end{abstract}

\section{Introduction }
Transparent electrode has important prospective applications in the fields of photodetectors, light-emitting diodes (LEDs), vertical cavity surface emitting lasers (VCSELs), and solar cells, et al \cite{NP12KE, AM11DSH}. As a novel two-dimensional (2-D) material, graphene has high conductivity and very high light transmission   within the range of medium infrared and visible light frequency \cite{AM11DSH, NL08XW, NN10SB}. Thus, it is regarded as an ideal transparent electrode material. However, the plasmon of graphene exhibits very high terahertz (THz) absorption when the carrier concentration of graphene is high within the THz or far-infrared frequency range \cite{NN11LJ, NL11LR, APL11BS, PRB11JH, PRB08ZZZ, APL13MM, JPCM13NMRP, PCCP13YXZ, JAP14XH, NN14XC}. Therefore, graphene with high carrier concentration in THz frequency range is regarded as an absorption medium instead of a transparent electrode.

The frequency of the THz range is mainly the spectrum within 0.1-10 THz. Given its special properties, the spectrum of the THz range has broad applications in communication, medical imaging, radar detection, nondestructive tests, and so on \cite{S09LYS, S12BE}.  The transparent electrode within the THz frequency range has important prospective applications in THz detectors, modulators,  VCSELs, and phase shifters \cite{S09LYS, S12BE, APL11BS, NL11LR, NN11LJ, IEEE14GKS, LSA15LW, OE13CSY, OE12RM,  OE13YW, OL14CSY}. Numerous THz transparent electrodes have been suggested, including  two-dimensional electron gas (2DEG), ion-gel, two-dimensional arrays of metallic square holes, graphene with low carrier density, and indium-tin-oxide (ITO) nanomaterials. However, Broadband ultra high transmission THz transparent electrodes is still very desirable.

Recently, another 2-D material, namely, monolayer MoS$_{2}$, has attracted much research attention. monolayer MoS$_{2}$ has high mobility ($\sim$410 cm$^{2}$V$^{-1}$s$^{-1}$) and excellent mechanical properties \cite{NN12QHW, PRL10KFM, NL10AS, NN13OLS, AN11ZY, NL12HSL, AM12WC, NL13MB, JPCC07TL, AN13YC, NL12JP, OL14CJ}. However, its optical property is not the same as that of graphene. The monolayer MoS$_{2}$ band gap is not zero, which has a high absorption   within the visible-light frequency range. This material can be used as a high-efficiency photoelectric detector material instead of an ideal transparent electrode material within the visible light frequency \cite{NN12QHW, PRL10KFM, NL10AS, JPCC07TL, NN13OLS, AN11ZY, NL12HSL, AM12WC, NL13MB, PRL14CHL, A15XDY}. So, the question about whether monolayer MoS$_{2}$ can be used to make transparent electrodes within the THz frequency range is interesting.

This study investigated the monolayer MoS$_{2}$ absorption and transmission at the THz wave band and compared its results with those of graphene and 2DEG. The results show that monolayer MoS$_{2}$ has lower THz absorption than graphene by one to three grades. When the carrier concentration was  10$^{12}$ cm$^{-2}$, the THz absorption   of monolayer MoS$_{2}$ was less than 2.4\% and its transmission   was much higher than that of 2DEG. For example, when the carrier concentration was  $8.4\times10^{12}$ cm$^{-2}$, the relative transmission amplitude of InAs 2DEG was approximately 55\%, and the relative transmission amplitude of monolayer MoS$_{2}$ could reach 95\%. As a representative case, we calculated the field-effect tubular structure formed by monolayer MoS$_{2}$-insulation layer-graphene, which would allow the THz of graphene to reach saturation under low voltage. The THz absorption by monolayer MoS$_{2}$ was approximately 5\% at the maximum.

\section{Model and Theory}

The experimental results show that the dielectric constant of monolayer MoS$_{2}$ within the THz frequency range can be characterized by the Drude model $\varepsilon(\omega)=\varepsilon_{\infty}-\frac{\omega_{p}^{2}}{\omega^{2}+i\Gamma\omega}$, where  $\omega_{p}^{2}=\frac{Ne^{2}}{2\varepsilon_{0}m^{*}}$  is the plasma frequency \cite{A15XDY}. Within the  THz frequency range, the conductivity of graphene can be expressed as \cite{EPL13JTL}
$\sigma_{g}=\frac{e^{2}}{\pi \hbar }\frac{|\epsilon_{F}|}{\hbar \Gamma -i\hbar \omega},$
where $\hbar$ is the reduced Planck constant,  $\hbar \Gamma=2.5$ meV is the relaxation rate, $\epsilon_{F}$ is the Fermi
level position with respect to the Dirac point, and $\omega$ is the angular frequency
 of the incident THz radiation. The  permittivity of graphene can given by  $\varepsilon_{g}(\omega)=1+i\frac{\sigma_{v}}{\omega \varepsilon_{0}}=1+i\frac{\sigma_{g}}{\omega \varepsilon_{0}d_{g}}$, where $\varepsilon_{0}$ is the vacuum permittivity, $\sigma_{v}$ is the conductivity of bulk materials, and $d_{g}=0.34$ nm is the thickness of graphene.

The standard transfer-matrix method was used for the calculation  \cite{EPL13JTL,JAP14JTL}.
In the \emph{l}th layer, the electric field of the TE mode light with incident angle $\theta_{i}$ is given by
\begin{equation}\mathbf{E}_{l}(z,y)=\left[  A_{l}e^{ik_{lz}%
\left(  z-z_{l}\right)  }+B_{l}e^{-ik_{lz}\left(  z-z_{l}\right)  }\right]
e^{ik_{ly}y}\mathbf{e}_{x}, \label{TMM:a1}\end{equation}
and the magnetic field of the TM mode  is given by
\begin{equation}\mathbf{H}_{l}(z,y)=\left[  A_{l}e^{ik_{lz}%
\left(  z-z_{l}\right)  }+B_{l}e^{-ik_{lz}\left(  z-z_{l}\right)  }\right]
e^{ik_{ly}y}\mathbf{e}_{x},\end{equation}
where $k_{l}=k_{lr}+ik_{li}$ is the wave vector of the light, $\mathbf{e}_{x}$ is the unit vectors in the x direction, and $z_{l}$ is the position of the \emph{l}th layer in the z direction.

The electric fields of TE mode or the magnetic fields of TM mode in the (\emph{l}+1)th layer are related to the incident fields by the transfer matrix utilizing the boundary condition \cite{EPL13JTL,JAP14JTL}.
Thus, we can obtain  the absorbance of \emph{l}th layer $\mathcal{A}_{l}$  using the Poynting vector $\textbf{S}=\textbf{E}\times\textbf{H}$ \cite{EPL13JTL,JAP14JTL}
\begin{equation}
\mathcal{A}_{l}=[{S}_{(l-1)i}+{S}_{(l+1)i}-{S}_{(l-1)o}-{S}_{(l+1)o}]/{S}_{0i},
\end{equation}
where  ${S}_{(l-1)i}$ and ${S}_{(l-1)o}$ [${S}_{(l+1)i}$ and ${S}_{(l+1)o}$] are the incident  and  outgoing Poynting vectors in (\emph{l}-1)th [(\emph{l}+1)th] layer, respectively, ${S}_{0i}$ is the incident Poynting vectors in air.

\section{NUMERICAL RESULTS}

\begin{figure}[H]
\includegraphics[width=0.95\columnwidth,clip]{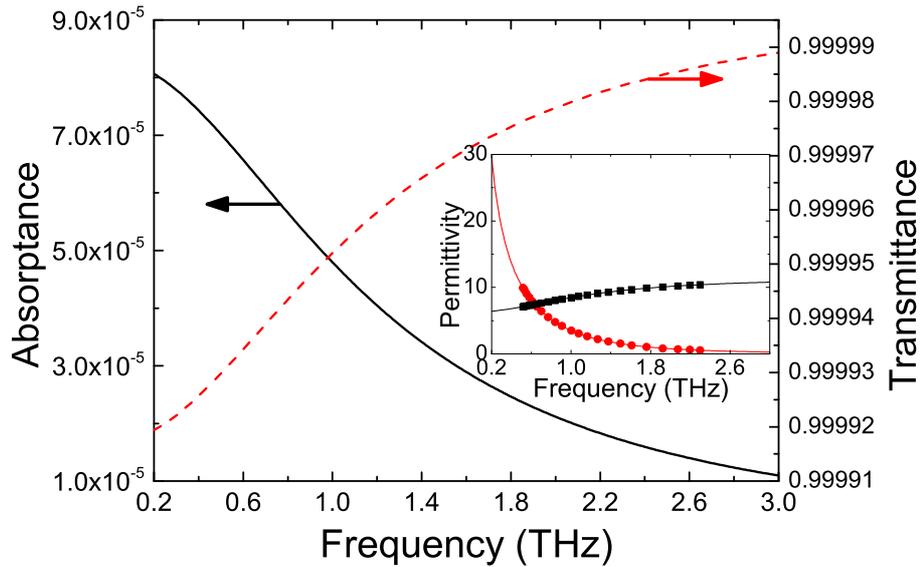}
\caption{(Color online)  The absorption   (black solid line) and transmission   (red dashed line) of monolayer MoS$_{2}$
as they change with the frequency. The inset shows that the dielectric constant of monolayer MoS$_{2}$ changes
with the wave length, in which the black blocks and red dots are the real and imaginary parts respectively
 of the dielectric constant measured in the experiment \cite{A15XDY}. The black solid  and red dashed lines represent the
 results of fitting. }
\label{fig1}%
\end{figure}

We calculated the absorption and transmission   of monolayer MoS$_{2}$ when THz  $\omega_{p}=16.77$ THz  at $N\approx3.3\times10^{9}$ cm$^{-2}$  (same as the experimental measurement result  \cite{A15XDY}).  The calculation result is indicated in Figure 1. The absorption rate of monolayer MoS$_{2}$ was less than 10$^{-4}$, its reflection rate was even smaller, and the transmission rate was larger than 0.9999. The main reason for this trend is that the effective quality of the monolayer MoS$_{2}$ carrier is relatively high, which causes low ion frequency and reduces the absorption of THz waves. Simultaneously, the real and imaginary parts of the monolayer MoS$_{2}$ dielectric constant are relatively small (see inset of Figure 1). The thickness of monolayer MoS$_{2}$ was very small at approximately 0.65 nm, and thus the monolayer MoS$_{2}$ reflection rate  and absorption is small. Consequently, monolayer MoS$_{2}$ has a very high transmission rate within the THz range.

\begin{figure}[H]
\includegraphics[width=0.95\columnwidth,clip]{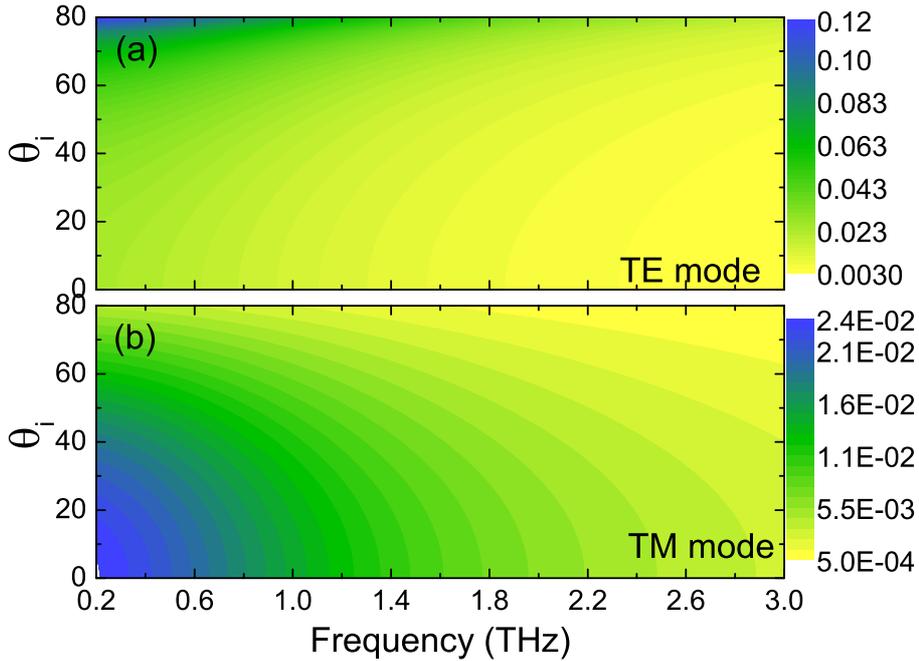}
\caption{(Color online)Contour plots
of the equivalent loss of  monolayer MoS$_{2}$ as a function
of the frequency  and the incident angles for
the (a) TE and (b) TM modes.}
\label{fig2}%
\end{figure}

Even for oblique incidence, the monolayer MoS$_{2}$ has a very high transmission   within the THz range. The equivalent loss (i.e., 1-T, where T is the transmission  ) of  monolayer MoS$_{2}$ with carrier concentration  10$^{12}$ cm$^{-2}$ as a function of the frequency  and the incident angles for the TE and TM modes are shown in Fig. 2.  When the incidence angle $\theta_{i}$ increases, the equivalent loss of TE mode is enhanced and the equivalent loss of TM mode is reduced. But even for TE mode with $\theta_{i}=80 ^{\circ}$, the equivalent loss is only about 12\%.

\begin{figure}[H]
\includegraphics[width=0.95\columnwidth,clip]{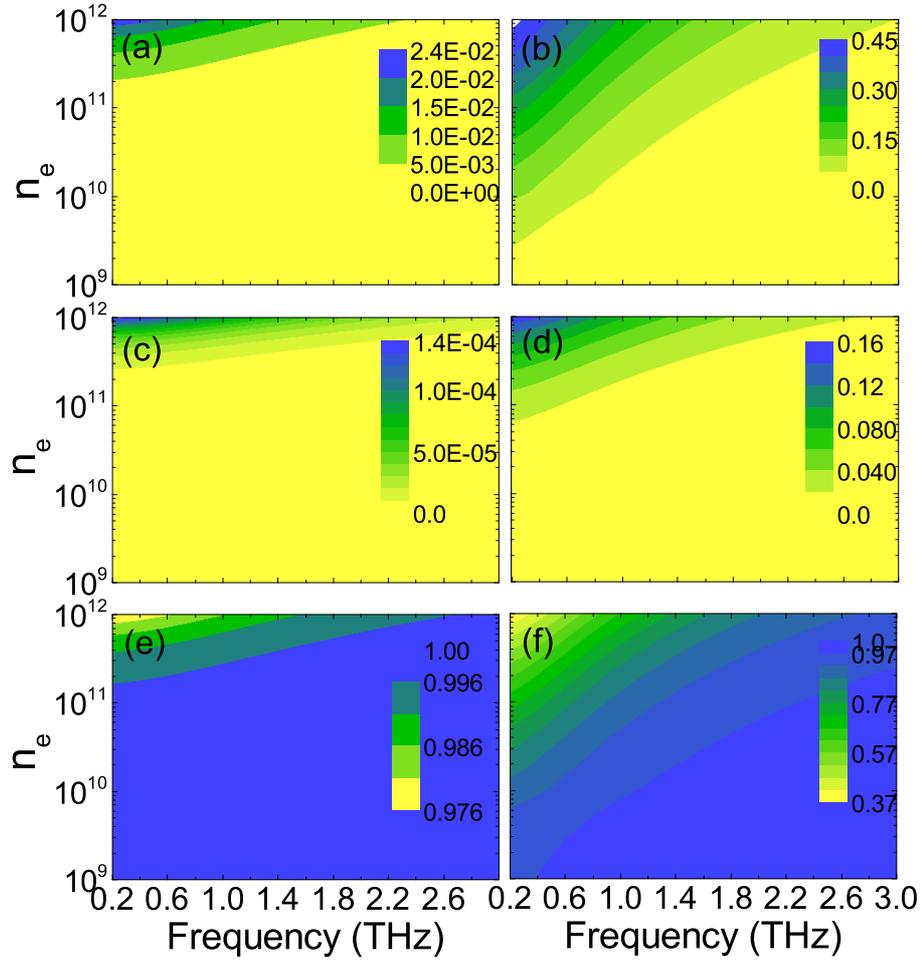}
\caption{(Color online)contour line diagram of the monolayer MoS$_{2}$ absorption   (a),
 reflection   (c), and transmission   (e), which all change with the frequency and carrier concentration.
  The contour line diagram of graphene absorption (b), reflection   (d),
  and transmission   (f), which all change with the frequency and carrier concentration. }
\label{fig3}%
\end{figure}

In a real THz device, the electrode usually has high carrier concentration to reduce electrode resistance and increase grid voltage. We studied the THz absorption   of monolayer MoS$_{2}$, as well as the reflection and transmission  under different carrier concentrations. As a point of comparison, we also provided the absorption, reflection, and transmission  of graphene. The calculation results are presented in Figure 3. When carrier concentration is increased, the ion frequency and THz absorption   of monolayer MoS$_{2}$ are increased, whereas the transmission   is reduced. According to the Drude modes, the THz absorption   of monolayer MoS$_{2}$ will decrease with the THz wave frequency. Even when the carrier concentration is at $N=10^{12}$ cm$^{-2}$, the wave absorption   of monolayer MoS$_{2}$ at a frequency of 0.2 THz is only 2.4\%. Thus, the reflection   can be ignored, and the transmission   is larger than 97\%. In comparison, when the carrier concentration is at $N=10^{12}$ cm$^{-2}$, the wave absorption at the frequency of 0.2 THz by graphene  is as high as 46\%. Meanwhile, the reflection   is 16\%, and the transmission   is only 37\%.

\begin{figure}[H]
\includegraphics[width=0.95\columnwidth,clip]{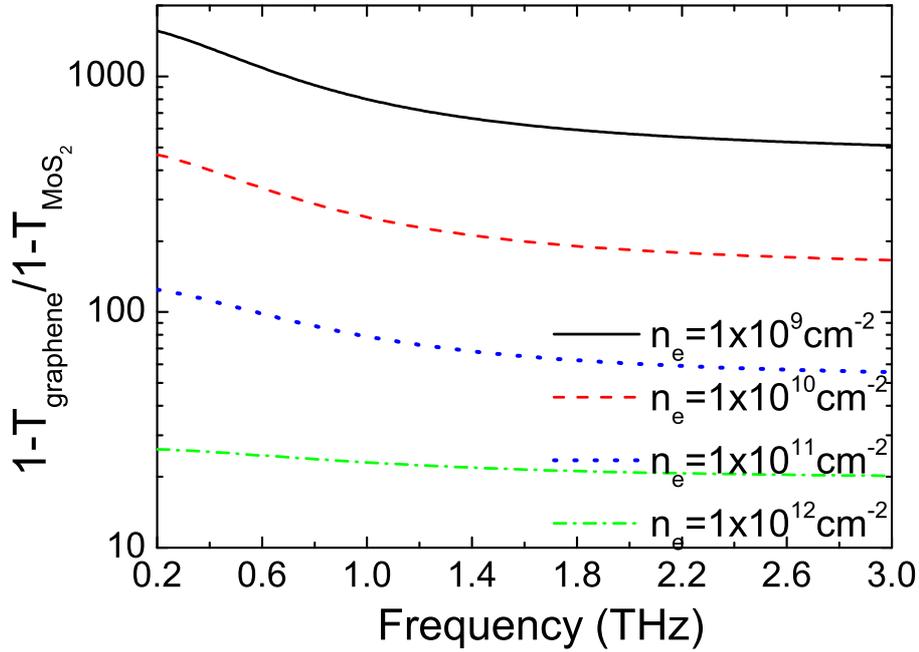}
\caption{(Color online)The ratio of equivalent loss between monolayer MoS$_{2}$ and graphene under different carrier concentrations.}
\label{fig4}%
\end{figure}

To compare the THz transmission   in monolayer MoS$_{2}$ and graphene, Figure 4 shows the ratio of equivalent loss between graphene and monolayer MoS$_{2}$ (the sum of reflection and absorption). When the carrier concentration is low, such as $N=10^{9}$ cm$^{-2}$  and  $f=0.2$ THz, the equivalent loss of monolayer MoS$_{2}$ is smaller than  $4\times10^{-5}$, which is three grades smaller that that of graphene. When the carrier concentration is increased, the ratio between graphene and monolayer MoS$_{2}$ is reduced. The main reason for this outcome is that even when the carrier concentration is low, such as  $N=10^{9}$ cm$^{-2}$, the equivalent loss of graphene is still high, which is approximately 4\%. When the carrier concentration is increased, the increase in equivalent loss of graphene is limited, particularly when the carrier concentration is high. Thus, the equivalent loss of graphene is close to saturation.

\begin{figure}[H]
\includegraphics[width=0.95\columnwidth,clip]{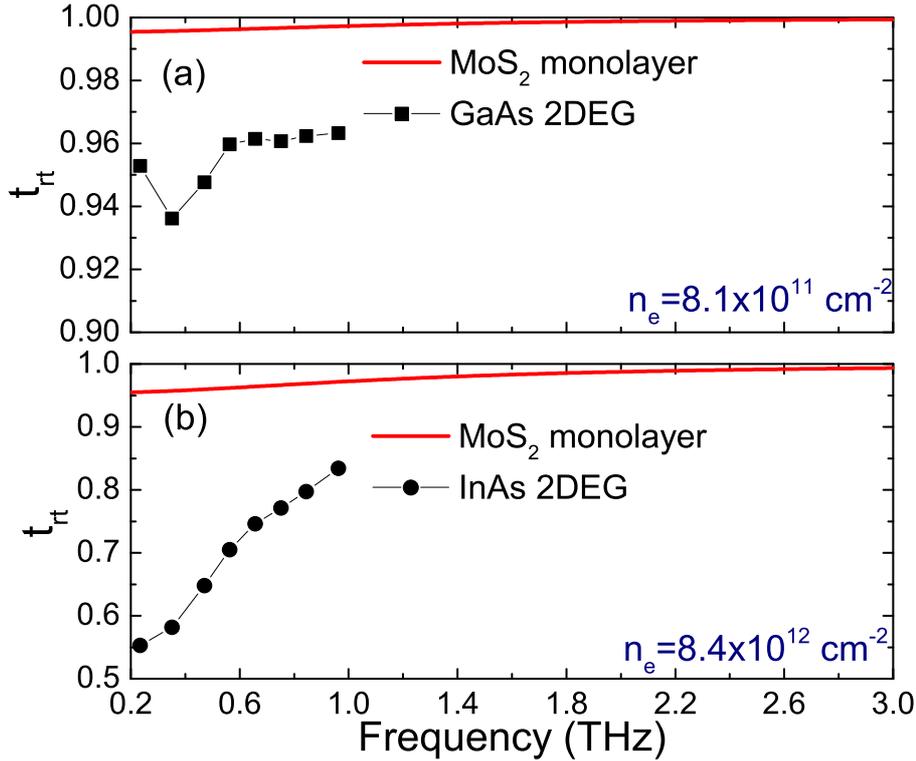}
\caption{(Color online) the relative transmission amplitude of monolayer MoS$_{2}$, and compared with the experimental results of
 (a) GaAs 2DEG, and (b) InAs 2DEG \cite{APL06NAK} under different carrier concentrations.  }
\label{fig3}%
\end{figure}

Under actual conditions, the transparent electrode normally remains on the surface of the base. We studied the effect of monolayer MoS$_{2}$ on the transmission of samples when it is on the base surface and compared it with the existing experimental results for 2DEG \cite{APL06NAK}. Thus, the relative transmission amplitude can be defined as  $t_{rt}=1+n_{sub}+Z_{0}\sigma_{s}$. When  $N=8.1\times10^{11}$ cm$^{-2}$, the relative transmission amplitude of monolayer MoS$_{2}$ is still larger than 0.995, whereas in InAs 2DEG, this value is  about 0.96 [see Fig. 5 (a)].  When  $N=8.4\times10^{12}$ cm$^{-2}$, the relative transmission amplitude of monolayer MoS$_{2}$ is still larger than 0.95, whereas in InAs 2DEG, this value is only approximately 0.55 [see Fig. 5 (b)]. Given that the change in actual transmission   is the norm of the relative transmission amplitude, the effect of monolayer MoS$_{2}$ on the transmission   is even smaller compared with 2DEG.

\begin{figure}[H]
\includegraphics[width=0.95\columnwidth,clip]{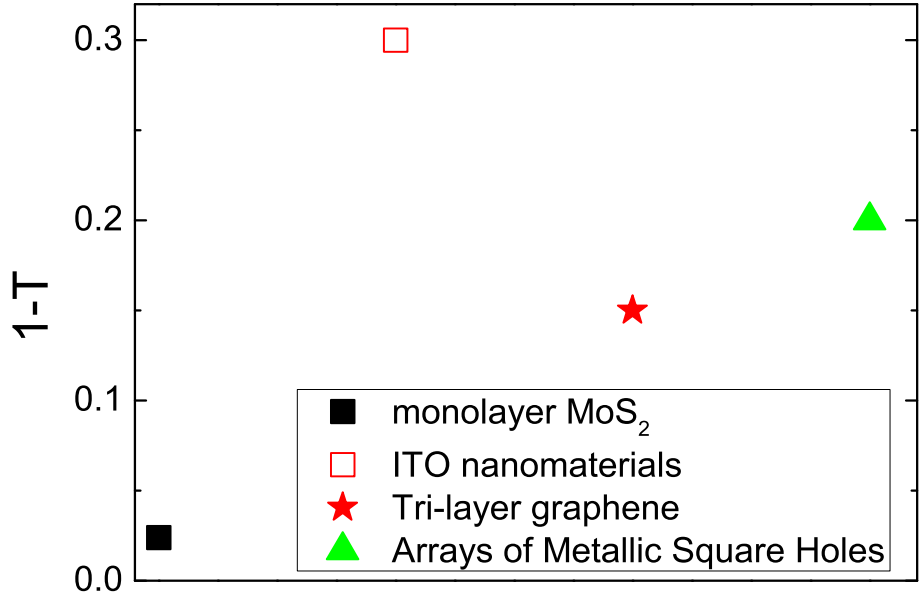}
\caption{(Color online)The equivalent loss of  monolayer MoS$_{2}$, two-dimensional arrays of metallic square holes
 \cite{IEEE14GKS}, Tri-layer graphene \cite{LSA15LW}, and ITO nanomaterials \cite{OE13CSY}.}
\label{fig6}%
\end{figure}

Although the details of the transparent electrodes are different,   we can still make a comparison with other THz transparent electrodes [Fig. 6]. The maximum equivalent loss of  monolayer MoS$_{2}$ with $N=10^{12}$ cm$^{-2}$ is about  2.4\%, which is much small than that of the ITO nanomaterials \cite{OE13CSY} and tri-layer graphene \cite{LSA15LW}. The maximum transmission of  two-dimensional arrays of metallic square holes  can reach nearly 100\%, but the  average equivalent loss within broadband frequency regions  of the metallic square holes can reach 20\% \cite{IEEE14GKS}.

\begin{figure}[H]
\includegraphics[width=0.95\columnwidth,clip]{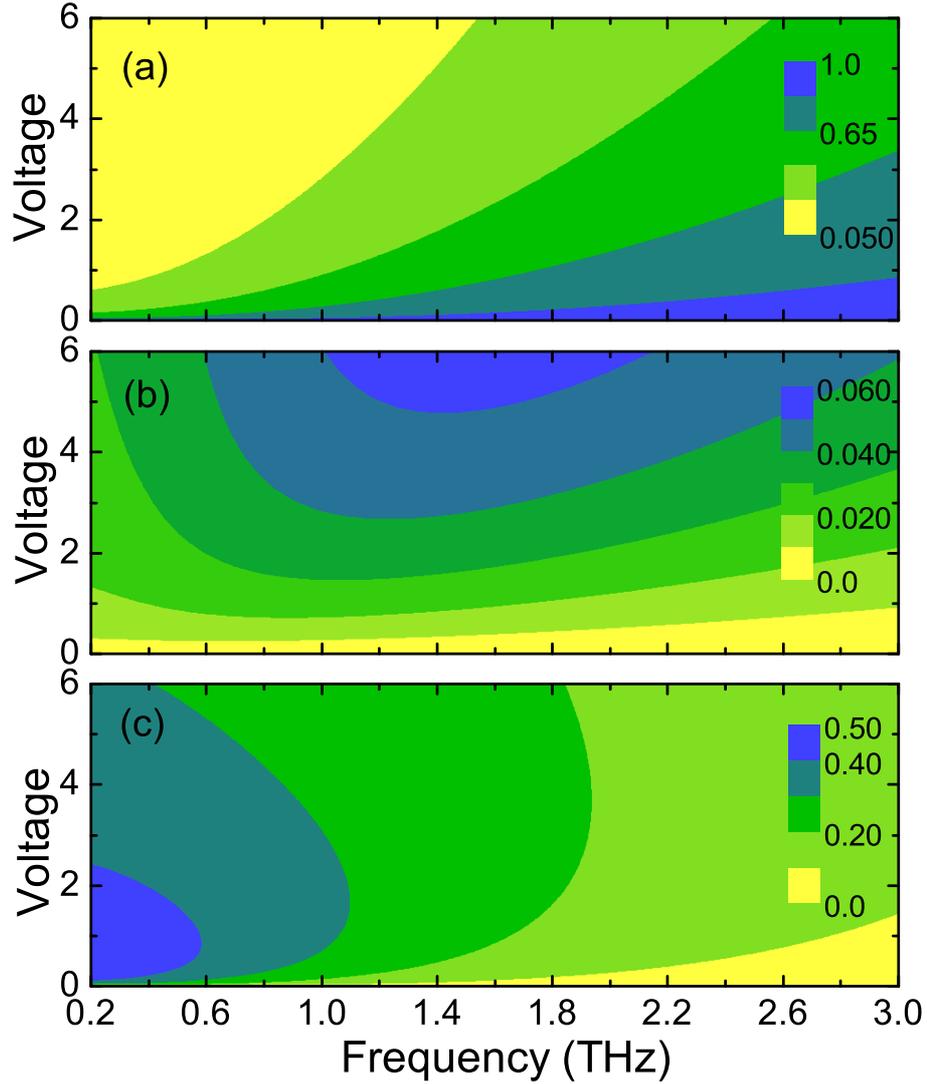}
\caption{(Color online)the contour line diagram of the (a) total transmission, (b) monolayer MoS$_{2}$ absorption,
and (c) graphene absorption   as a function of the frequency and grid voltage in the sandwich structure.}
\label{fig7}%
\end{figure}

To reveal the properties of monolayer MoS$_{2}$ in real devices, we studied the THz absorption by the field effect formed by monolayer MoS$_{2}$-insulation layer-graphene, as well as the regulation effect of the grid voltage between monolayer MoS$_{2}$ and graphene on THz. A similar structure was used to make the modem or detector for THz.
When the thickness of the insulation layer is smaller, the grid voltage required for modulation will be smaller. But if the thickness of the insulation layer is excessively small, the tunneling current will rapidly increase. The breakdown electric field of the SiO$_{2}$ film is relatively strong and can reach 20 MV/cm \cite{APL07CS}. In the calculation, the insulation takes 3 nm of the SiO$_{2}$ layer. The initial Fermi energy of monolayer MoS$_{2}$ and graphene is 0, and the calculation result is illustrated in Figure 7. When the grid voltage is 0, the absorption of THz by monolayer MoS$_{2}$ and graphene is very small, and the THz wave is almost completely broken down. When the grid voltage increases, the carrier concentrations of monolayer MoS$_{2}$ and graphene are increased and the THz transmission   is reduced. When the grid voltage is at 6 V, its transmission   is only 5\%. However, the maximum absorption   of monolayer MoS$_{2}$ is only 5.4\%. Most of the voltage is absorbed or reflected by graphene. In this structure, the absorption of monolayer MoS$_{2}$ and graphene is not at the lowest frequency and the maximum grid voltage because most THz waves are reflected by graphene instead of being absorbed under very high carrier concentration.

\section{Conclusions}
The study demonstrated that the THz absorption by monolayer MoS$_{2}$ is very small over a broadband frequency range even under high carrier concentration and large incident angle. Its equivalent loss is one to three grades smaller than that of graphene. The transmission of monolayer MoS$_{2}$  is much larger than that of traditional GaAs or InAs electron gas. We studied the field-effect tubular structure of monolayer MoS$_{2}$-insulation layer-graphene, which allows the THz absorption of graphene to reach saturation even under low voltage even while the THz absorption of monolayer MoS$_{2}$ is only approximately 5\% at the maximum. Therefore, monolayer MoS$_{2}$ can be used to make transparent electrodes within the THz frequency range. This development has important prospective applications in optoelectronic devices within the THz frequency range.

\end{document}